%% file: main.tex
\documentclass[10pt,journal,compsoc]{IEEEtran}

\usepackage[nocompress]{cite}
\usepackage{amsmath,amssymb}
\usepackage{booktabs}
\usepackage{array}
\usepackage{graphicx}
\usepackage{multirow}
\usepackage{algorithm}
\usepackage{algpseudocode}
\usepackage{xcolor}
\usepackage{stfloats}
\usepackage[hidelinks]{hyperref}

\newtheorem{proposition}{Proposition}

\title{Reducing Fixed-Batch Client Idleness in Many-Client Edge Split Learning}

\author{Mohammad~Kohankhaki, Valentin~Rentschler, Carsten~Kamp, Sigrid~Brell-Cokcan, and~Anke~Schmeink%
\IEEEcompsocitemizethanks{\IEEEcompsocthanksitem M. Kohankhaki, V. Rentschler, and A. Schmeink are with the Chair of Information Theory and Data Analytics, RWTH Aachen University, Kopernikusstr. 16, 52074 Aachen, Germany.\protect\\
Corresponding author: Mohammad Kohankhaki.\protect\\
E-mail: mohammad.kohankhaki@inda.rwth-aachen.de
\IEEEcompsocthanksitem C. Kamp and S. Brell-Cokcan are with the Chair of Individualized Production in Architecture, RWTH Aachen University, Campus-Boulevard 30, 52074 Aachen, Germany.
\IEEEcompsocthanksitem M. Kohankhaki, C. Kamp, S. Brell-Cokcan, and A. Schmeink are also with the Cluster of Excellence CARE, TU Dresden and RWTH Aachen, Germany.}}
\hypersetup{pdftitle={Reducing Fixed-Batch Client Idleness in Many-Client Edge Split Learning},pdfauthor={Mohammad Kohankhaki, Valentin Rentschler, Carsten Kamp, Sigrid Brell-Cokcan, Anke Schmeink}}

\begin{document}
\relpenalty=10000
\binoppenalty=10000
\IEEEtitleabstractindextext{%
\begin{abstract}
Parallel Split Learning with Global Sampling (GPSL) trains a split model on a fixed-size batch drawn from the data of all clients. When the client population is much larger than the batch, most clients contribute no data in a round, and additional server capacity does not change this. We propose Global Clustered Parallel Split Learning (GCPSL), which assigns clients to stable clusters, runs one GPSL workload per cluster concurrently, and periodically fuses the split-model replicas. We derive the expected direct data participation under the batch sampler, which separates repeated sampling of the same clients from idling at the fusion barrier. With 256 clients, more clusters raise participation but reduce the client diversity seen by each workload. In three paired runs on four H100 GPUs, GCPSL raises direct data participation from 22\% to 57--60\%. With random or size-balanced clusters, it reaches 85\% CIFAR-10 validation accuracy in 5.7 minutes, 1.6 times faster than GPSL with its batch sharded across the same GPUs. It matches per-step synchronized data-parallel GPSL at 85\% and reaches 88\% in about 10 instead of 19.5 minutes, without per-step global gradient or BatchNorm synchronization. Concurrent execution is 3.11 times faster than sequential execution of the same workloads.
\end{abstract}

\begin{IEEEkeywords}
Split learning, parallel split learning, edge computing, distributed training, client participation, non-IID data
\end{IEEEkeywords}}

\IEEEpubid{\begin{minipage}{\textwidth}\footnotesize\raggedright This work has been submitted to the IEEE for possible publication. Copyright may be transferred without notice, after which this version may no longer be accessible.\end{minipage}}
\maketitle
\IEEEdisplaynontitleabstractindextext

\input{sections/01_intro_related}
\input{sections/02_system_method}

\input{sections/03_methodology_results}

\input{sections/04_discussion_conclusion}

\section*{Acknowledgments}
Funded by the Deutsche Forschungsgemeinschaft (DFG, German Research Foundation) under Germany's Excellence Strategy -- EXC 3115 -- 533767731.
The authors used Claude (Anthropic) for drafting assistance and language editing throughout the manuscript. The authors reviewed and revised all text and take full responsibility for the content.

\input{sections/05_references}
\end{document}

%% file: sections/01_intro_related.tex
\IEEEraisesectionheading{\section{Introduction}\label{sec:introduction}}
\IEEEPARstart{M}{anaged} edge systems may train models across cameras, sensors, vehicles, and institutional devices while keeping raw data local. Split learning supports this setting by dividing a neural network between clients and a server. Clients compute the early layers and send cut-layer activations rather than raw examples. The server completes the forward and backward pass and returns cut gradients \cite{gupta2018distributed}. Parallel split learning lets several clients contribute to one training step, but the composition of that step matters when client data are not independent and identically distributed (non-IID) \cite{jeon2020privacypsl,cai2022efficient,hsu2020federated}.

Parallel Split Learning with Global Sampling (GPSL) addresses this statistical problem by drawing a fixed-size batch from the distributed client data \cite{kohankhaki2025gpsl}. The batch size does not grow with the number of clients, which bounds both per-step work and direct data participation. If $K$ clients share a batch of $B$ examples, no more than $B$ clients can supply data in that round. When $K$ is much larger than $B$, most clients perform no data-dependent client-side computation in that round, even when the server has spare capacity.

\begin{figure}[t]
\centering
\includegraphics[width=\columnwidth]{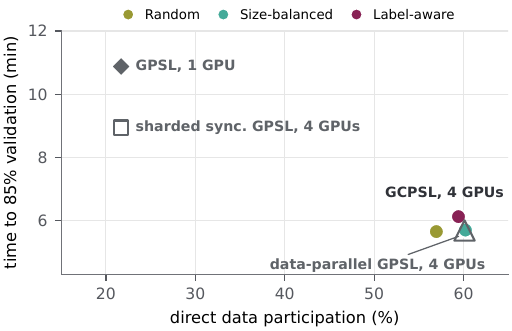}
\caption{Measured operating points on CIFAR-10, as means over three paired runs. GPSL uses one GPU. Sharded synchronous GPSL, data-parallel GPSL, and the three GCPSL assignments use four. Lower time and higher participation are better.}
\label{fig:teaser}
\end{figure}

\begin{figure*}[t]
\centering
\includegraphics[width=\textwidth]{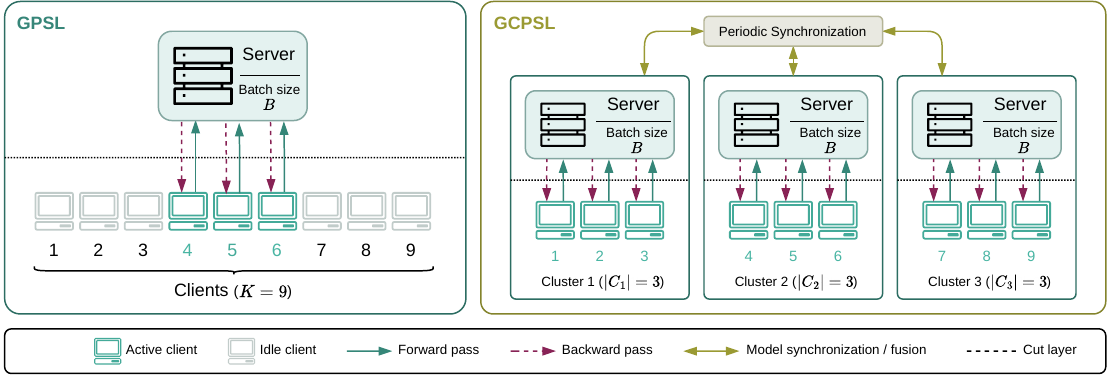}
\caption{Execution organization. GPSL uses one global workload with batch size $B$, so only clients represented in that batch supply data. GCPSL assigns clients to stable clusters and runs one GPSL workload with batch size $B$ in each cluster. The cluster workloads execute concurrently, and their split-model replicas are fused at periodic barriers. Clients that supply no data in a round still receive the synchronized client-model update.}
\label{fig:overview}
\end{figure*}

Additional server GPUs can be used in three standard ways, and each has a cost. A larger batch lets more clients contribute per round but increases memory use and changes the optimization regime \cite{keskar2017largebatch,shallue2019measuring}. Sharding the batch across GPUs shortens each step but leaves participation unchanged. Data-parallel training with $B$ examples per GPU raises participation (Fig.~\ref{fig:teaser}) but synchronizes gradients, and in our baseline also server normalization statistics, across all GPUs at every step. We propose Global Clustered Parallel Split Learning (GCPSL), which assigns clients to stable clusters, runs one GPSL workload with the original batch per cluster, and fuses the split-model replicas only at epoch boundaries (Fig.~\ref{fig:overview}). GCPSL builds on our GPSL design~\cite{kohankhaki2025gpsl}. It reuses the GPSL sampler, and our evaluation reuses the GPSL client-partition protocol. The clustered execution, the participation analysis, and all experiments in this paper are new.

This paper makes four contributions.
\begin{itemize}
  \item We introduce GCPSL, which combines stable client clustering, cluster-local GPSL, and periodic split-model fusion. Clusters train concurrently without moving raw client data and exchange model state only at fusion boundaries.
  \item We derive the expected direct data participation under the GPSL sampler. It separates repeated sampling of the same clients from idling at the fusion barrier and, for large client datasets, bounds the expected fraction of a cluster that one batch reaches.
  \item We measure how cluster count and assignment information trade participation against accuracy, comparing random, size-balanced, and label-aware assignment.
  \item We implement GCPSL with PyTorch Distributed and compare it with five GPSL baselines on up to four H100 GPUs. With random or size-balanced clusters, GCPSL reaches 85\% validation accuracy 1.6 times faster than sharded synchronous GPSL, matches data-parallel GPSL at 85\%, and reaches 88\% in about half its time. Concurrent execution achieves 78\% of the ideal four-way speedup over sequential execution of the same workloads.
\end{itemize}

Section~\ref{sec:related} reviews related work. Sections~\ref{sec:system} and~\ref{sec:method} define the system model and method, Sections~\ref{sec:methodology} and~\ref{sec:results} present the evaluation, Section~\ref{sec:discussion} discusses the operating region and limitations, and Section~\ref{sec:conclusion} concludes.

\section{Related Work}
\label{sec:related}

\subsection{Split and Parallel Split Learning}
Split learning keeps raw samples with their owners while moving intermediate activations and gradients across a client-server boundary \cite{gupta2018distributed}. Parallel variants let multiple clients interact with server-side model segments at the same time \cite{jeon2020privacypsl,cai2022efficient}. SplitFed combines split execution with federated aggregation \cite{thapa2022splitfed}. Cluster-based parallel split learning (CPSL) trains the clients of one cluster in parallel and the clusters one after another, and it optimizes clustering and radio resources for latency \cite{wu2023splitwireless}. GCPSL inverts this organization. It samples GPSL batches within each cluster and runs the clusters concurrently. Efficient parallel split learning reduces training latency over resource-constrained wireless edge networks \cite{lin2024epsl}, and distribution-aware clustered split learning shows that grouping clients changes statistical behavior \cite{arafat2026dcsl}. These systems do not study the participation limit that arises when GPSL keeps one fixed global batch for a much larger client population.

GPSL draws each batch from the global distributed data instead of concatenating independent local mini-batches \cite{kohankhaki2025gpsl}. This construction improves batch representativeness under non-IID data, but the fixed batch limits how many clients can supply examples in one round. GCPSL keeps GPSL sampling within each cluster and periodically fuses the resulting model replicas.

\subsection{Participation and Workload Organization}
Federated optimization studies data heterogeneity, participant selection, and time to accuracy in decentralized learning \cite{mcmahan2017communication,li2020fedprox}. Oort selects useful clients under system variation \cite{lai2021oort}. TiFL groups clients by training performance, FedScale provides infrastructure for large client populations, and HACCS clusters clients under heterogeneous resources \cite{chai2020tifl,lai2022fedscale,wolfrath2022haccs}. Clustered federated learning and IP-FL use groups for distribution structure and personalized objectives \cite{sattler2020clustered,khan2025ipfl}. These methods aggregate client-computed model updates. GCPSL instead executes a split forward and backward path for each active example, so client grouping also controls server workload length, activation traffic, and the state that must be fused.

Non-IID benchmarks show why client grouping cannot be chosen from load alone \cite{hsu2020federated,li2022niidbench}. A balanced number of samples may still yield a poor class mixture, while a statistically balanced cluster can have a longer schedule. GCPSL therefore evaluates three metadata budgets: no data-derived metadata, one sample count per client, and sample counts with class histograms. To our knowledge, this participation and composition trade-off has not been studied for fixed-batch split learning.

\subsection{Concurrent and Distributed Training}
Distributed training systems reduce iteration time through synchronization, placement, and communication scheduling. PipeDream pipelines model partitions, ByteScheduler reorders communication, MCR-DL selects communication mechanisms, and QSync coordinates heterogeneous accelerators \cite{narayanan2019pipedream,peng2019bytescheduler,anthony2023mcrdl,zhao2024qsync}. Edge deployments add limits on computation, memory, and communication \cite{satyanarayanan2017edge,zhou2019edgeintelligence,shi2020communication}. Larger batches and compressed synchronization offer additional capacity and communication choices \cite{shallue2019measuring,lin2018deep}. Local SGD averages worker replicas only periodically and can generalize better than large mini-batch training \cite{stich2019localsgd,lin2020localsgd}, and hierarchical federated learning aggregates at edge servers before a global aggregation \cite{liu2020hierarchical}. GCPSL applies periodic averaging to split-learning replicas, each trained by one cluster-local GPSL workload on its own execution slot.

%% file: sections/02_system_method.tex
\section{System Model}
\label{sec:system}

\subsection{Entities and Assumptions}
We consider $K$ stable clients in a managed edge system. Client $k$ owns a dataset $\mathcal D_k$, and raw examples remain on that client. Together, the client datasets form the training dataset $\mathcal D$, with $|\mathcal D|=\sum_k|\mathcal D_k|$. The clients may have different sample counts and label distributions. A coordinator can assign clients to server execution slots and can collect the metadata required by the chosen assignment rule. Client membership does not change during training.

A split model consists of a client segment $f_{\theta_c}$ and a server segment $g_{\theta_s}$. A participating client evaluates $z_k=f_{\theta_c}(x_k)$ on its batch examples $x_k$ and sends the activations $z_k$ to the server. The server completes the forward pass, computes the loss $\mathcal L$, and returns the cut-layer gradient $\nabla_{z_k}\mathcal L$. GCPSL maintains one client-segment state $\theta_{c,n}$ and one server-segment state $\theta_{s,n}$ for each cluster $C_n$. The server offers $N$ execution slots, but this model does not assume a particular processor or network mapping.

An execution slot holds one server replica and processes one cluster workload. Slots may share a host or occupy different hosts. Clients send activations to their cluster's server replica and receive cut gradients. At fusion, the coordinator exchanges model parameters and server normalization state across slots. The coordinator is trusted to create assignments and perform fusion.

A \emph{local round} is one GPSL update within a cluster. An \emph{epoch} ends after every example assigned to that cluster has been scheduled once. Clusters with different sample counts may need different numbers of local rounds. The system epoch ends at a fusion barrier after all clusters finish. Shorter workloads wait there without taking additional updates.

We distinguish two forms of client participation. A client has \emph{direct data participation} when at least one of its examples enters the current batch. It has \emph{model-update participation} when it applies the shared client-segment update. A client without direct data participation performs no client-side forward or backward computation in that round. We call this \emph{inactivity}. GPSL and GCPSL can update a client replica even when that client supplies no example. Direct data participation and model-update participation are therefore different quantities.

\subsection{Communication and Assignment}
During local training, a client exchanges activations and cut gradients only with its assigned cluster's server replica. Active client-side gradients are combined within that cluster and applied to every client replica there. At a fusion barrier, the coordinator combines client- and server-model parameters and server normalization state across clusters, then broadcasts the result.

Random clustering uses no data-derived metadata, size-balanced clustering uses one sample count per client, and label-aware clustering uses sample counts and class histograms (Section~\ref{sec:method}). The metadata determine the fixed assignments before training and are not sent with each batch.

GCPSL keeps client assignments fixed and exchanges model state across clusters only at fusion boundaries. It requires no per-step communication between clusters.

\subsection{Fixed-Batch Participation Limit}
GPSL fixes a global batch size $B$ independently of $K$ \cite{kohankhaki2025gpsl}. Let $b_{k,t}\in\mathbb Z_{\geq 0}$ be the number of examples requested from client $k$ in round $t$, with $\sum_{k=1}^{K} b_{k,t}\leq B$. The direct participant set of round $t$ is
\begin{equation}
 A_t=\{k\mid b_{k,t}>0\}.
 \label{eq:gpsl_batch}
\end{equation}
Every client in $A_t$ supplies at least one example, so $|A_t|\leq B$. The fraction of clients with no direct data participation satisfies
\begin{equation}
 \rho_t=1-\frac{|A_t|}{K}
 \geq \max\!\left(0,1-\frac{B}{K}\right).
 \label{eq:gpsl_idle_bound}
\end{equation}
For example, $K=256$ gives a lower bound of 50\% at $B=128$ and 75\% at $B=64$. This limit follows from the fixed batch rather than from client or server speed. GCPSL reduces the client population served by each workload while preserving its batch size.

\section{GCPSL Method}
\label{sec:method}

\subsection{Clustered Workloads}
GCPSL partitions the $K$ clients into a set $\mathcal C=\{C_1,\ldots,C_N\}$ of $N$ disjoint clusters, and each cluster runs one GPSL workload with batch size $B$ (Fig.~\ref{fig:overview}). Cluster $C_n$ owns $M_n=\sum_{k\in C_n}|\mathcal D_k|$ training examples, and the mean cluster sample count is $\bar M=|\mathcal D|/N$. In local round $t$, cluster $n$ requests at most $B$ examples. Let $A_{n,t}=\{k\in C_n\mid b_{k,t}>0\}$ be its direct participant set and $\rho_{n,t}=1-|A_{n,t}|/|C_n|$ its inactivity. Then
\begin{equation}
 \rho_{n,t}\geq\max\!\left(0,1-\frac{B}{|C_n|}\right).
 \label{eq:cluster_idle_bound}
\end{equation}
With about $K/N$ clients per cluster, increasing $N$ can therefore raise direct data participation, but each GPSL sampler then sees fewer clients and may see less label diversity. The assignment rule controls this participation and composition trade-off.

\emph{Random clustering} permutes the clients and assigns them round-robin to clusters of nearly equal cardinality. It uses no data-derived client metadata and has $O(K)$ assignment cost after the permutation.

\emph{Size-balanced clustering} uses one sample count $|\mathcal D_k|$ per client. It sorts clients by decreasing sample count and places each client in the cluster with the smallest current sample count among those that stay within $1.25\,\bar M$. Ties are broken deterministically. Construction costs $O(K\log K+KN)$.

\emph{Label-aware clustering} uses sample counts and class histograms. Let $p_n$ be the sample-weighted class distribution in cluster $n$ and $p$ the global class distribution. Starting from a random round-robin assignment, the rule repeatedly applies the single-client move that most reduces
\begin{equation}
 J(\mathcal C)=
 \frac{\sum_n |M_n-\bar M|}{|\mathcal D|}
 +\lambda\sum_n\frac{M_n}{|\mathcal D|}\,d_{\mathrm{JS}}(p_n,p).
 \label{eq:cluster_objective}
\end{equation}
Here $d_{\mathrm{JS}}$ is the Jensen--Shannon distance, the square root of the Jensen--Shannon divergence with the natural logarithm, and $\lambda=12$ weights class alignment against sample balance. The first term balances cluster sample counts, and the second aligns each cluster's class distribution with the global distribution. Moves keep every cluster nonempty and $M_n\leq1.25\,\bar M$. The search stops when no move improves $J$ or after 3000 moves. Each move evaluates at most $K(N-1)$ candidates, and clustering runs once before training.

Algorithm~\ref{alg:clustering} summarizes the three rules, which all produce a fixed map before training. Stable membership lets each execution slot retain its sampler, optimizer, client replicas, and server replica across local rounds.

\begin{algorithm}[t]
\caption{Stable client clustering}
\label{alg:clustering}
\begin{algorithmic}[1]
\Require clients $\{1,\ldots,K\}$, the client metadata required by rule $h$, clusters $N$, rule $h$
\If{$h=\text{random clustering}$}
  \State permute clients
  \State assign them round-robin
\ElsIf{$h=\text{size-balanced clustering}$}
  \State sort clients by decreasing sample count
  \ForAll{clients in sorted order}
    \State assign the client to the lightest allowed cluster
  \EndFor
\Else
  \State assign permuted clients round-robin
  \Repeat
    \State evaluate allowed moves with Eq.~\eqref{eq:cluster_objective}
    \State apply the best improving move, if any
  \Until{no allowed move improves $J$ or 3000 moves are applied}
\EndIf
\State \Return fixed clusters $\{C_n\}_{n=1}^{N}$
\end{algorithmic}
\end{algorithm}

\begin{algorithm}[t]
\caption{GCPSL training}
\label{alg:gcpsl}
\begin{algorithmic}[1]
\Require datasets $\{\mathcal D_k\}$, $N$, $B$, epochs $E$, rule $h$
\State $\{C_n\}\gets\Call{StableCluster}{\{1,\ldots,K\},N,h}$
\State initialize all split-model replicas equally
\For{$e=1,\ldots,E$}
  \ForAll{$n=1,\ldots,N$ \textbf{ in parallel}}
    \For{$t=1,\ldots,\lceil M_n/B\rceil$}
      \State draw GPSL batch $t$ from $C_n$
      \State active clients send cut activations
      \State server $n$ returns matching cut gradients
      \State update the server replica
      \State combine active client gradients
      \State update every client replica in $C_n$
    \EndFor
  \EndFor
  \State fuse parameters and normalization state
  \State broadcast the fused split model
\EndFor
\State \Return fused split model
\end{algorithmic}
\end{algorithm}

\subsection{Cluster-Local GPSL}
Each cluster runs GPSL over its assigned clients. In a local round, the active clients compute cut-layer activations and the cluster's server replica returns the corresponding cut gradients. The active clients' gradients are summed under the batch-mean loss, which weights each client by its number of examples in the batch. Every client replica in $C_n$ applies the result, including replicas whose owners supplied no data in that round. Thus all client replicas in one cluster share a common client model.

Client-segment batch normalization (BatchNorm) layers~\cite{ioffe2015batchnorm} are replaced with group normalization (GroupNorm)~\cite{wu2018group} to avoid divergent client-local running statistics. The server segment retains BatchNorm. Each server replica maintains its running mean, running variance, and batch counter between fusion boundaries. These quantities form the server normalization state that is fused with the model parameters.

\subsection{Fusion and Broadcast}
The evaluated method fuses the cluster replicas after every epoch. Let $q_n=M_n/|\mathcal D|$ be the sample weight of cluster $n$. The fused parameters are
\begin{equation}
 \bar\theta_c=\sum_{n=1}^{N}q_n\theta_{c,n},
 \qquad
 \bar\theta_s=\sum_{n=1}^{N}q_n\theta_{s,n}.
 \label{eq:fusion}
\end{equation}
Server running means and variances use the same weights, so each cluster contributes in proportion to its data. The BatchNorm batch counter takes the maximum over clusters. The fused parameters and normalization state are broadcast to all replicas before the next epoch, and validation and test evaluation use this common model. Optimizer states, including momentum, are not averaged. Each cluster resumes with its own optimizer state after receiving the common parameters.

\subsection{Schedule and Cost}
Cluster $n$ takes $r_n=\lceil M_n/B\rceil$ local updates per epoch, and stable clusters keep $r_n$ fixed across epochs. Concurrent execution completes the local part of an epoch after $T=\max_n r_n$ update slots. A shorter cluster waits at the fusion barrier and takes no padded optimizer steps. Sequential execution of the same workloads requires $\sum_n r_n$ slots. The schedule speedup of concurrent over sequential execution is
\begin{equation}
 S=\frac{\sum_n r_n}{T}\leq N.
 \label{eq:schedule_speedup}
\end{equation}
GCPSL maintains $N$ cluster-level split-model states, with one server replica per cluster and synchronized client-segment replicas within each cluster, while keeping the memory demand of one $B$-sample workload on each execution slot. If the fused state contains $P$ bytes, an epoch boundary communicates $O(NP)$ logical state. Activation and cut-gradient payload remains batch dependent within each workload. Algorithm~\ref{alg:gcpsl} summarizes the two synchronization levels: client-gradient combination in every cluster-local round and model fusion across clusters at the epoch boundary.

\subsection{Direct Data Participation Under the Sampler}
\label{sec:expected-participation}
We account for concurrent execution in system phases. Phase $t\in\{1,\ldots,T\}$ groups local round $t$ of every cluster with $r_n\geq t$, while the remaining clusters wait at the barrier. Phases are an accounting device: clusters do not synchronize between local rounds, only at the fusion barrier. With $A_{n,t}=\emptyset$ for $t>r_n$, the direct participant set of phase $t$ is $A_t=\bigcup_n A_{n,t}$, which reduces to Eq.~\eqref{eq:gpsl_batch} for $N=1$. Each workload has at most $B$ direct participants, so
\begin{equation}
 \rho_t
 =1-\frac{|A_t|}{K}
 \geq\max\!\left(0,1-\frac{NB}{K}\right).
 \label{eq:system_idle_bound}
\end{equation}

This bound counts each batch example as a different client. Two effects keep inactivity above it: the sampler can draw several examples from one client, and clients of shorter clusters idle at the barrier. To separate them, let $\pi$ be the fraction of the $KT$ client--phase slots of an epoch with direct data participation, and let $\mu$ be the fraction in which the client's cluster is active, with $\mu=1$ for equal cluster lengths. In an active slot the client applies the cluster's client-model update, so $\mu$ is the model-update participation rate of Section~\ref{sec:system}. The within-round rate $\bar\pi$ is the fraction of active slots with direct data participation. By construction,
\begin{equation}
 \pi=\mu\,\bar\pi,
 \qquad
 \mu=\frac{1}{KT}\sum_{n=1}^{N}|C_n|\,r_n.
 \label{eq:participation-factorization}
\end{equation}

At the start of each epoch, the sampler of cluster $n$ lists one entry per training example, labeled with its owner, shuffles the list uniformly at random, and splits it into $r_n$ consecutive batches whose sizes differ by at most one. Let $B_{n,t}=\sum_{k\in C_n}b_{k,t}\leq B$ be the size of batch $t$.

\begin{proposition}\label{prop:participation}
Fix the clusters, and let $k\in C_n$ and $t\leq r_n$.

(a) Client $k$ participates directly in round $t$ with probability
\begin{equation}
 \Pr[k\in A_{n,t}]
 =1-\binom{M_n-|\mathcal D_k|}{B_{n,t}}\bigg/\binom{M_n}{B_{n,t}},
 \label{eq:expected-participation}
\end{equation}
where $\binom{a}{b}=0$ for $b>a$. The expected within-round rate $\mathbb E\bar\pi$ is the mean of these probabilities over all active client--phase slots.

(b) Let $\phi_k=|\mathcal D_k|/M_n$ be the data share of client $k$ in its cluster. For a batch of size $B_{n,t}=B\geq2$, as all sample counts in $C_n$ grow with fixed shares,
\begin{equation}
 \begin{split}
 \frac{\mathbb E|A_{n,t}|}{|C_n|}
 &\to
 \frac{1}{|C_n|}\sum_{k\in C_n}\bigl[1-(1-\phi_k)^B\bigr]\\
 &\leq 1-\Bigl(1-\frac{1}{|C_n|}\Bigr)^{B},
 \end{split}
 \label{eq:participation-bound}
\end{equation}
with equality if and only if all clients in $C_n$ hold equal shares.
\end{proposition}

\begin{IEEEproof}
(a) The uniform shuffle makes every set of $B_{n,t}$ entries equally likely to form batch $t$. Exactly $\binom{M_n-|\mathcal D_k|}{B_{n,t}}$ of the $\binom{M_n}{B_{n,t}}$ sets contain no example of client $k$. The statement on $\mathbb E\bar\pi$ follows from linearity of expectation, which requires no independence between clients or rounds.
(b) The ratio of binomial coefficients equals $\prod_{j=0}^{B-1}(M_n-|\mathcal D_k|-j)/(M_n-j)$, which tends to $(1-\phi_k)^B$ as $M_n$ grows with $\phi_k$ fixed. The function $\psi(\phi)=1-(1-\phi)^B$ is strictly concave on $[0,1]$ for $B\geq2$, and $\sum_{k\in C_n}\phi_k=1$. Jensen's inequality gives $\sum_{k\in C_n}\psi(\phi_k)/|C_n|\leq \psi(1/|C_n|)$, with equality if and only if all $\phi_k$ are equal.
\end{IEEEproof}

For $B=64$, the equal-share value in Eq.~\eqref{eq:participation-bound} is 22.2\% for one workload over 256 clients and 63.5\% for clusters of 64 clients. Choosing $NB=K$ therefore removes the counting bound in Eq.~\eqref{eq:system_idle_bound}, yet in the large-pool limit more than a third of the active client--phase slots remain without direct data participation. Size-balanced clustering nearly equalizes the cluster totals $M_n$ and hence the lengths $r_n$, which raises $\mu$ in Eq.~\eqref{eq:participation-factorization}. It does not equalize client shares within a cluster, so in the large-pool limit $\bar\pi$ stays below the equal-share value. Section~\ref{sec:fixed-assignment} tests both statements.

\subsection{Distributed Implementation}
\label{sec:implementation}
We implement GCPSL in PyTorch~\cite{paszke2019pytorch} with PyTorch Distributed~\cite{li2020pytorchdistributed} and the NVIDIA Collective Communications Library (NCCL)~\cite{nvidia_nccl} and run one process per execution slot. Each process holds the client-segment and server-segment replicas of one cluster together with its sampler and optimizer. At the start of an epoch, rank 0 builds the batch schedule of every cluster from the run seed and the epoch index and broadcasts it. Each rank then runs its $r_n$ local rounds, and a rank with fewer rounds waits at the barrier without taking an optimizer step. At fusion, the ranks reduce their sample-weighted parameters and server normalization buffers to rank 0, which broadcasts the fused state. All ranks then compare SHA-256 hashes of their model state, so every epoch starts from verified identical replicas. Optimizer states stay local. The synchronous baselines use the same code with a per-step gradient all-reduce and globally synchronized server BatchNorm.

Code and per-run results are available at \url{https://github.com/mohkoh19/gcpsl}.

%% file: sections/03_methodology_results.tex
\input{figures/generated/tpds_target_rows.tex}
\input{figures/generated/tpds_timing_rows.tex}
\input{figures/generated/tpds_sync_rows.tex}
\section{Experimental Methodology}
\label{sec:methodology}
We evaluate GCPSL in two settings. Schedule-accounted studies explore the design space: cluster count, assignment rule, fusion interval, batch size, and dataset. Real-execution studies measure wall-clock time on four H100 GPUs and compare GCPSL with other ways of using the same GPUs.

\subsection{Model and Data}
We evaluate CIFAR-10 and CIFAR-100 with a ResNet-18 model for $32\times32$ inputs, split after its first residual stage \cite{krizhevsky2009cifar,he2016resnet}. The client segment holds the stem convolution and the first residual stage with GroupNorm, and the server segment holds the remaining three stages and the classifier with BatchNorm (Section~\ref{sec:method}). Training uses random crops, horizontal flips, dataset normalization, cross-entropy with label smoothing 0.1, and stochastic gradient descent (SGD) with momentum 0.9, Nesterov acceleration, weight decay $5\times10^{-4}$, and a constant learning rate. Unless noted otherwise, the learning rate is 0.01 and training lasts 100 epochs.

The client partition follows the extended-Dirichlet protocol of GPSL \cite{kohankhaki2025gpsl}. Each of the $K=256$ logical clients is assigned support over two classes, and samples within the selected classes are allocated with concentration $\alpha=3$. The primary CIFAR-10 setting uses $B=64$, for which Eq.~\eqref{eq:gpsl_idle_bound} gives a 75\% inactivity bound. We also study CIFAR-10 at $B=128$ and use $B=128$ for CIFAR-100.

\subsection{Schedule-Accounted Studies}
A single-process implementation trains all cluster workloads one after another with the sampling, update, and fusion rules of GCPSL. Its learning outcomes are measured, but its time is counted in idealized schedule rounds: an epoch costs the local rounds of the longest cluster, as in Eq.~\eqref{eq:schedule_speedup}. One schedule round is one system phase of Section~\ref{sec:expected-participation}. $R_q$ denotes the rounds needed to reach $q$\% test accuracy. These studies train on all 50,000 training examples and process five million examples per run. Processed samples separate a shorter schedule from training on more examples. Batch deviation is the $\ell_1$ distance between a batch's class distribution and the global training distribution.

The studies sweep $N\in\{2,4,8,16\}$ for label-aware and size-balanced GCPSL and evaluate random GCPSL at $N=8$. Fusion controls compare one-epoch fusion, fusion every 100 steps and at epoch ends, and no fusion. Capacity controls train one GPSL workload with $B\in\{256,512\}$ and learning rate 0.04, the best of 0.005 to 0.04 in CIFAR-10 test accuracy for one seed. Within each cluster-count comparison, the data partition, model split, optimization settings, and example budget are fixed, so changing $N$ changes only the number and composition of cluster workloads. Results are means and standard deviations over three runs. A separate memory control runs this implementation for ten epochs on one NVIDIA RTX A4000 and records peak GPU memory in use.

\subsection{Real-Execution Study}
The real-execution study trains on CIFAR-10 with $K=256$, $N=4$, and $B=64$ per workload on one node with four NVIDIA H100 GPUs. All runs use PyTorch 2.6.0 and CUDA 12.4 in one container image, with FP32 arithmetic and TF32 matrix multiplication. Each GPU runs one process of the implementation in Section~\ref{sec:implementation}. Cluster-local activations and cut gradients stay within the owning process, and only fusion and synchronization collectives cross GPUs. The 50,000 training examples are split into 45,000 training and 5,000 validation examples before client partitioning. The 10,000 test examples are used only for final reporting. Every regime runs with seeds 40, 72, and 313, and runs with the same seed are paired across regimes.

We compare random, size-balanced, and label-aware GCPSL under the same allocation, per-workload batch, and epoch fusion. A sequential control runs the same four label-aware workloads in turn on the same allocation, which isolates the effect of concurrent execution. Five GPSL baselines use the GPUs differently.
\begin{itemize}
  \item The \emph{single-worker} baseline is the original GPSL: one workload with $B=64$ on one GPU.
  \item The \emph{larger-batch} baseline runs one workload with $B=256$ on one GPU.
  \item The \emph{sharded synchronous} baseline splits one batch of $B=64$ across four GPUs and trains one global model.
  \item The \emph{data-parallel} baseline draws one GPSL batch of 256 examples, gives 64 to each GPU, and trains one global model.
  \item The \emph{repartitioned periodic} baseline runs four workloads with epoch fusion as in GCPSL but reassigns clients to workloads at every fusion.
\end{itemize}
The sharded synchronous and data-parallel baselines synchronize gradients and server BatchNorm statistics across all four GPUs at every step. For the larger-batch and data-parallel baselines, a 30-epoch validation-only sweep over learning rates 0.005, 0.01, 0.02, 0.04, and 0.08 selected 0.02. All other runs use 0.01.

The primary endpoint is $t_{85}$, the wall-clock time until the first epoch whose validation accuracy reaches 85\%. The times $t_{80}$ and $t_{88}$ to 80\% and 88\% are secondary endpoints. All three targets were fixed before the runs. The timer starts before the first epoch and covers setup, client clustering, training, collectives, barrier waiting, and every validation pass up to the crossing. GPU-worker-minutes multiply $t_{85}$ by the number of allocated GPUs. Test accuracy is reported at epoch 100 and at the best-validation checkpoint. Direct data participation is $\pi$ in Eq.~\eqref{eq:participation-factorization}, averaged over all epochs. Logical application payload counts each client, server, and fusion tensor transfer once, up to the target. Tables report means and sample standard deviations over the three paired runs.

\section{Results}
\label{sec:results}

\subsection{Cluster Count Trades Participation for Local Diversity}
\begin{figure*}[t]
\centering
\includegraphics[width=0.98\textwidth]{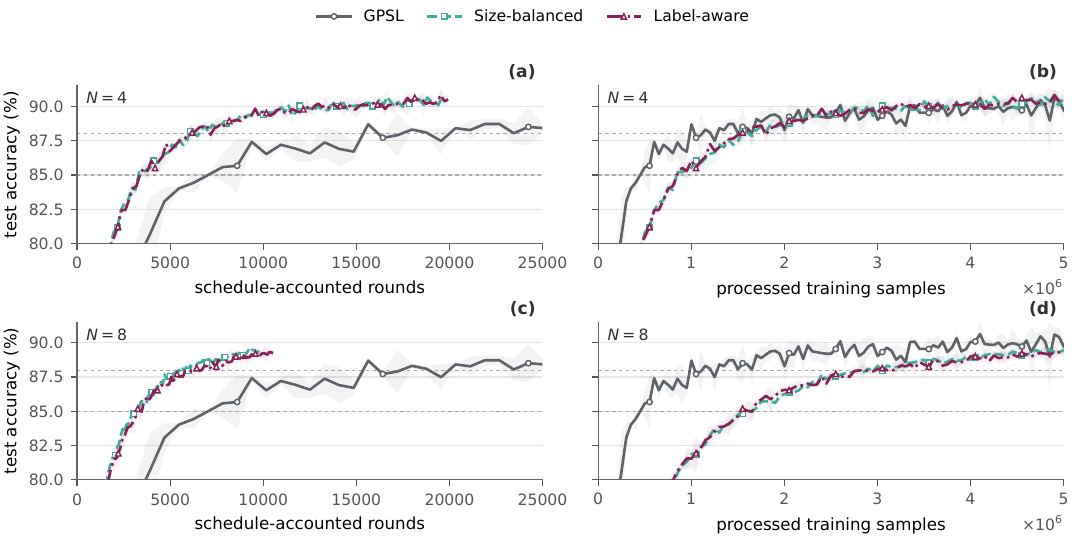}
\caption{CIFAR-10 schedule-accounted study with $K=256$ and $B=64$. The top row compares GPSL with size-balanced and label-aware GCPSL at $N=4$, and the bottom row uses $N=8$. The left column plots test accuracy against schedule rounds, and the right column against processed samples. The y-axes show the 80--91.5\% target region. Shaded regions show standard deviations over three runs.}
\label{fig:s2_accounted_time}
\end{figure*}

GCPSL reaches the CIFAR-10 targets in fewer schedule rounds than GPSL at $B=64$. Figure~\ref{fig:s2_accounted_time}(a) and (c) show the schedule-accounted trajectories for $N=4$ and $N=8$. Per processed example, GCPSL learns no faster than GPSL, and at $N=8$ it learns more slowly, as panels (b) and (d) show. Its gain comes from shorter schedules.

Table~\ref{tab:fixed-b} gives the cluster-count results. At $N=4$, label-aware GCPSL reduces $R_{85}$ by 50.0\% relative to GPSL and finishes 0.78 percentage points higher. Size-balanced GCPSL reduces $R_{85}$ by 51.7\% with similar final accuracy. Size-balanced GCPSL gives the lowest $R_{85}$ at $N=16$ and the lowest $R_{88}$ at $N=8$, using only sample-count metadata.

\begin{table*}[t]
\caption{Cluster-count study on CIFAR-10 with $B=64$.}
\label{tab:fixed-b}
\centering
\small
\setlength{\tabcolsep}{5.0pt}
\begin{tabular*}{\textwidth}{@{\extracolsep{\fill}}llccccc@{}}
\toprule
Method & Assignment & $N$ & Acc. (\%) $\uparrow$ & $R_{85}$ $\downarrow$ & $R_{88}$ $\downarrow$ & Inactivity (\%) $\downarrow$ \\
\midrule
GPSL & global & -- & 89.73 $\pm$ 0.65 & 7.04 $\pm$ 0.78 & 13.55 $\pm$ 1.97 & 78.23 $\pm$ 0.03 \\
\midrule
\multirow{9}{*}{GCPSL} & \multirow{4}{*}{label-aware} & 2 & \textbf{92.05 $\pm$ 0.06} & 4.70 $\pm$ 0.39 & 7.44 $\pm$ 1.02 & 61.70 $\pm$ 0.08 \\
 & & 4 & 90.51 $\pm$ 0.31 & 3.52 $\pm$ 0.12 & 6.05 $\pm$ 0.09 & 39.29 $\pm$ 0.16 \\
 & & 8 & 89.25 $\pm$ 0.39 & 3.33 $\pm$ 0.16 & 5.90 $\pm$ 0.43 & 17.31 $\pm$ 0.16 \\
 & & 16 & 87.51 $\pm$ 0.19 & 3.05 $\pm$ 0.14 & -- & 4.34 $\pm$ 0.22 \\
\cmidrule(l){2-7}
 & \multirow{4}{*}{size-balanced} & 2 & 91.33 $\pm$ 0.36 & 4.69 $\pm$ 0.78 & 7.56 $\pm$ 0.45 & 61.70 $\pm$ 0.07 \\
 & & 4 & 90.69 $\pm$ 0.63 & 3.40 $\pm$ 0.11 & 6.27 $\pm$ 0.34 & 39.32 $\pm$ 0.12 \\
 & & 8 & 89.47 $\pm$ 0.23 & 3.20 $\pm$ 0.25 & \textbf{5.32 $\pm$ 0.06} & 17.25 $\pm$ 0.12 \\
 & & 16 & 86.85 $\pm$ 0.14 & \textbf{3.03 $\pm$ 0.10} & -- & \textbf{4.33 $\pm$ 0.19} \\
\cmidrule(l){2-7}
 & random & 8 & 89.37 $\pm$ 0.44 & 3.41 $\pm$ 0.14 & 6.06 $\pm$ 0.82 & 17.17 $\pm$ 0.05 \\
\bottomrule
\end{tabular*}
\par\smallskip
\begin{minipage}{0.96\textwidth}
\footnotesize
Acc. is final test accuracy. $R_q$ is in thousands of schedule rounds. Values are means $\pm$ standard deviations over three runs. A dash denotes an unreached target. Bold marks the best value in each column. Percentages and ratios in the text use unrounded values.
\end{minipage}
\end{table*}

Proposition~\ref{prop:participation}(b) gives a large-pool reference for the inactivity column. In the large-pool limit with $B=64$, inactivity is at least 77.8\% for GPSL over 256 clients and 60.5, 36.5, 13.1, and 1.6\% for $N=2$, 4, 8, and 16 clusters. GPSL measures 78.23\%, and label-aware GCPSL measures 61.70, 39.29, 17.31, and 4.34\%. The reference assumes equal client shares and cluster sizes, so the 0.4 to 4.2 point excess is expected.

The client-to-batch ratio controls the available gain. At $B=128$, GPSL reaches 85\% in $4.69\times10^3$ rounds and label-aware GCPSL at $N=4$ in $2.87\times10^3$ rounds, a 38.9\% reduction compared with 50.0\% at $B=64$, at a cost of 1.55 percentage points of final accuracy. The smaller gain agrees with the weaker fixed-batch bound at $B=128$ in Eq.~\eqref{eq:gpsl_idle_bound}.

\subsection{Assignment Information Changes Cluster Composition}
\begin{figure*}[t]
\centering
\includegraphics[width=0.99\textwidth]{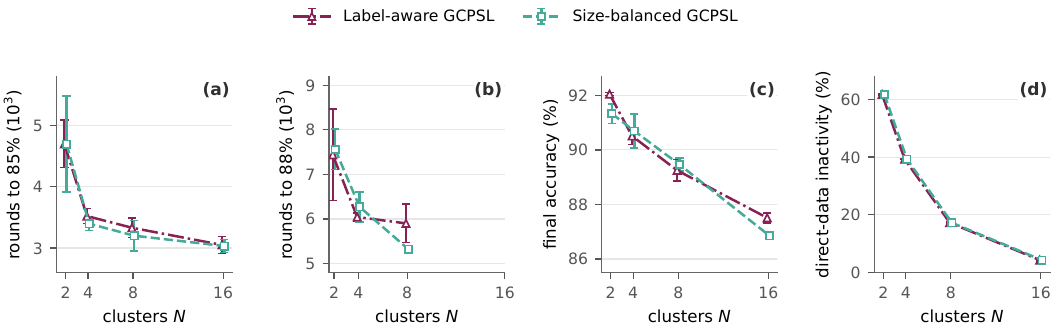}
\caption{Cluster-count trade-off for label-aware and size-balanced GCPSL on CIFAR-10. Panels (a) and (b) report rounds to the two targets, panel (c) final test accuracy, and panel (d) direct-data inactivity. Error bars show standard deviations over three runs. Horizontal offsets in panel (d) separate nearly coincident markers.}
\label{fig:s2_tradeoff}
\end{figure*}

Cluster count sets participation, and the assignment rule sets cluster composition. Figure~\ref{fig:s2_tradeoff}(d) shows almost identical inactivity for label-aware and size-balanced GCPSL at each $N$, yet their learning outcomes differ. Label-aware GCPSL gives the highest final accuracy at $N=2$, and size-balanced GCPSL reaches 85\% earlier for every $N\geq4$. At $N=16$, both rules lower inactivity below 5\% but lose 3.0 to 3.8 accuracy points relative to $N=4$.

\begin{table}[t]
\caption{CIFAR-100 schedule-accounted results.}
\label{tab:cifar100-boundary}
\centering
\small
\setlength{\tabcolsep}{2.7pt}
\input{figures/generated/v10_4_boundary_tabular.tex}
\par\smallskip
\begin{minipage}{0.98\columnwidth}
\footnotesize
$R_q$ is in thousands of schedule rounds, and Acc.\ is final test accuracy.
\end{minipage}
\end{table}

Label-aware clustering produces batches closest to the global class distribution. At $N=8$, its mean batch deviation is 0.30, compared with 0.42 for size-balanced and random GCPSL. On CIFAR-10 this composition advantage does not shorten the time to target. On CIFAR-100 it does: label-aware GCPSL at $N=4$ reaches 60\% in $4.05\times10^3$ rounds and finishes at 68.31\%, whereas size-balanced GCPSL needs $4.48\times10^3$ rounds and finishes at 66.77\% (Table~\ref{tab:cifar100-boundary}). With 100 classes and two classes per client, cluster composition matters more, and class histograms pay off.

\subsection{Fusion Is Required; Larger Batches Save Rounds}
Fusion couples the cluster replicas. Without fusion, label-aware GCPSL at $N=8$ finishes at 28.34\% (Table~\ref{tab:controls}(a)): the evaluation-time average of the unfused replicas is not a useful model. Fusion every 100 steps reaches 85\% slightly earlier than epoch fusion, but epoch fusion finishes 1.10 percentage points higher and reaches 88\% in every run. The real-execution study therefore uses epoch fusion. Executing the same eight workloads sequentially gives the same final accuracy but needs 7.5 times as many schedule rounds to reach 85\%.

\begin{table*}[t]
\caption{Fusion and payload controls.}
\label{tab:controls}
\centering
\begin{minipage}[t]{0.46\textwidth}
\centering
\small
\textit{(a) Fusion schedule, label-aware GCPSL at $N=8$}
\par\smallskip
\input{figures/generated/v10_4_fusion_tabular.tex}
\end{minipage}
\hfill
\begin{minipage}[t]{0.52\textwidth}
\centering
\small
\textit{(b) Analytical payload, single-process accounting}
\par\smallskip
\setlength{\tabcolsep}{4pt}
\input{figures/generated/v10_4_payload_tabular.tex}
\end{minipage}
\par\smallskip
\begin{minipage}{0.96\textwidth}
\footnotesize
$R_q$ is in thousands of schedule rounds. A dash denotes an unreached target. Without fusion, the cluster replicas are averaged only for evaluation. The 100-step interval counts local updates summed over all clusters, and sequential execution counts every local round. Payload in (b) uses decimal units and the single-process accounting, which counts each transfer at both endpoints and the client-model exchange with all $K$ clients in every phase, so it is comparable only within (b).
\end{minipage}
\end{table*}

In the single-process analytical accounting of Table~\ref{tab:controls}(b), payload to a target is the payload per phase times the number of phases. At $N=4$, GCPSL handles 28 to 30\% more payload per phase than GPSL because four workloads run at once, but it needs 50 to 52\% fewer phases to reach 85\% (Table~\ref{tab:controls}(b)). To reach 85\%, size-balanced GCPSL at $N=4$ handles 3.004~TB compared with 4.791~TB for GPSL. To reach 88\%, label-aware GCPSL at $N=4$ handles the least, 5.253~TB.

\begin{figure*}[t]
\centering
\includegraphics[width=\textwidth]{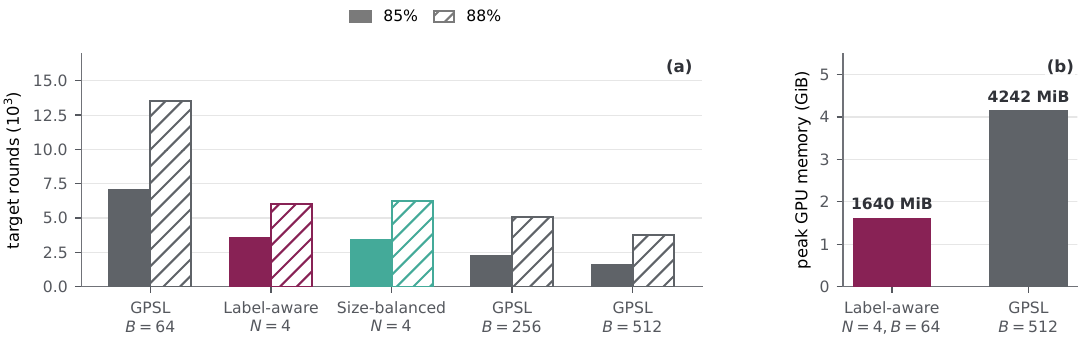}
\caption{Capacity boundary on CIFAR-10. (a) Rounds to 85\% and 88\% test accuracy for moderate-batch GPSL and GCPSL and for larger-batch GPSL, as means over three runs. (b) Peak GPU memory on one RTX A4000 for the process that holds all four label-aware GCPSL replicas ($N=4$, $B=64$) and for one GPSL workload with $B=512$.}
\label{fig:boundary}
\end{figure*}

In schedule rounds, a single larger batch is the stronger alternative when it fits in memory. GPSL with $B=512$ reaches 85\% in $1.57\times10^3$ rounds, compared with $3.40\times10^3$ for size-balanced GCPSL at $N=4$ (Fig.~\ref{fig:boundary}(a)). It finishes at 89.44\% on CIFAR-10, close to GPSL with $B=64$, but lowest on CIFAR-100, at 64.42\%. It also needs more memory. On one RTX A4000, one $B=512$ workload peaks at 4242~MiB. One $B=64$ workload peaks at 1080~MiB, and even the process that holds all four label-aware replicas at $N=4$ peaks at only 1640~MiB (Fig.~\ref{fig:boundary}(b)). In real execution on one H100, the larger-batch baseline with $B=256$ reaches 85\% at a similar time as single-worker GPSL and is the slowest method to 88\% (Section~\ref{sec:same-gpus}).

\subsection{Random and Size-Balanced GCPSL Match or Beat the Four-GPU Baselines at 85\% and 88\%}
\label{sec:same-gpus}
Table~\ref{tab:real-distributed} compares GCPSL with the five GPSL baselines, and Table~\ref{tab:target-sensitivity} adds the 80\% and 88\% targets. Every method reaches every target in every run.

\begin{table*}[t]
\caption{Real-execution results on CIFAR-10 over three paired runs.}
\label{tab:real-distributed}
\centering
\small
\input{figures/generated/real_results_mean_std_tabular.tex}
\par\smallskip
\begin{minipage}{0.96\textwidth}
\footnotesize
$t_{85}$ is the time to 85\% validation accuracy as mean $\pm$ sample standard deviation. Other columns are means. GPU-min is allocated GPU-worker time to the target. Acc.\ is epoch-100 test accuracy, and Best-val is the test accuracy of the best-validation checkpoint. Part.\ is direct data participation. Payload is logical application payload to the target in decimal TB.
\end{minipage}
\end{table*}

Sharding the GPSL batch across the four GPUs leaves direct data participation at 21.70\% and takes 8.95 minutes to reach 85\%. Random and size-balanced GCPSL raise participation to 56.96\% and 60.21\% and take 5.66 and 5.70 minutes, about 1.6 times faster. They reach 88\% in 10.07 and 10.20 minutes, compared with 16.85 minutes. Every GCPSL run reaches both targets before every sharded synchronous run.

\begin{table}[!t]
\caption{Time to validation targets in minutes (mean $\pm$ sample standard deviation over three paired runs).}
\label{tab:target-sensitivity}
\centering\small
\setlength{\tabcolsep}{5pt}
\begin{tabular}{@{}lrrr@{}}
\toprule
Method & 80\% & 85\% & 88\% \\
\midrule
\tpdsTargetRows
\bottomrule
\end{tabular}
\end{table}

Data-parallel GPSL is as fast as GCPSL at 85\%, with 5.69 minutes, and the fastest to 80\%. At 88\% the comparison reverses. Data-parallel GPSL needs 19.53 minutes and 61.7 epochs to reach 88\%, whereas GCPSL needs 10.07 to 11.30 minutes and 40.3 to 44.7 epochs, and every GCPSL run reaches 88\% before every data-parallel run. From 80\% to 88\%, the time grows by a factor of 6.7 for data-parallel GPSL and 7.0 for the larger-batch baseline, but by 3.0 to 3.2 for the three GCPSL assignments. These two baselines are the only methods whose updates use 256 examples and learning rate 0.02. The pattern agrees with reports that periodic model averaging generalizes better than training with large mini-batches \cite{lin2020localsgd}. Data-parallel GPSL also has the lowest epoch-100 test accuracy, 88.21\%, below each fixed assignment in all three paired runs.

The repartitioned periodic baseline matches GCPSL at both targets, with 5.84 and 10.31 minutes, and gives the highest final accuracy, 89.95\%. It reassigns clients to workloads at every fusion, whereas GCPSL keeps each client with one execution slot.

The single-GPU baselines use less GPU time. Single-worker GPSL reaches 85\% in 10.88 GPU-worker-minutes, compared with 22.81 for size-balanced GCPSL. The larger-batch baseline reaches the same participation as GCPSL on one GPU, 60.11\%, but takes 11.45 minutes. GCPSL thus spends additional GPU time to shorten wall-clock time at high participation.

\subsection{Concurrency Delivers Most of the Ideal Speedup}
\begin{figure*}[t]
\centering
\includegraphics[width=0.92\textwidth]{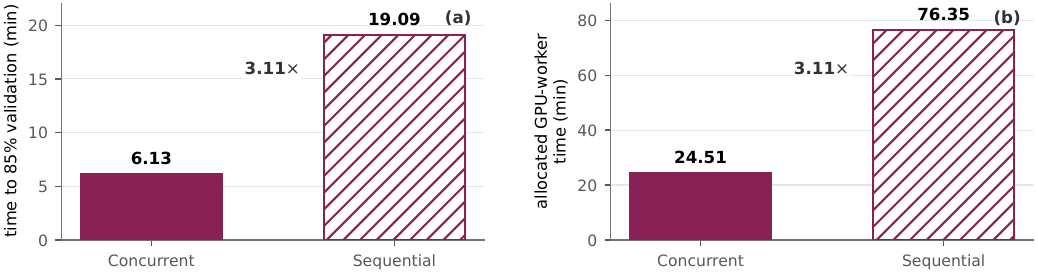}
\caption{The same four label-aware GCPSL workloads under concurrent and sequential schedules on four H100 GPUs, as means over three paired runs. (a) Wall-clock time to 85\% validation accuracy. (b) Allocated GPU-worker time to the same target. Standard deviations of the times in (a) appear in Table~\ref{tab:real-distributed}.}
\label{fig:concurrency}
\end{figure*}

Concurrent label-aware GCPSL reaches 85\% in $6.13\pm0.15$ minutes, and sequential execution of the same four workloads on the same GPUs takes $19.09\pm0.45$ minutes (Fig.~\ref{fig:concurrency}). The ratio of mean times, 3.11, is 77.9\% of the four-way ceiling in Eq.~\eqref{eq:schedule_speedup}. Both schedules use the same assignments, batches, fusion, participation, and logical payload. In every paired run they cross 85\% in the same epoch and finish with identical validation and test accuracy, so concurrency changes only the wall-clock time. Unequal cluster lengths, barrier waiting, fusion, and validation contribute to the remaining gap. The sequential schedule also uses 76.35 instead of 24.51 GPU-worker-minutes, because all four GPUs stay allocated while one workload runs.

\subsection{Assignment Changes Barrier Padding, Not Within-Round Participation}
\label{sec:fixed-assignment}
Size-balanced clustering raises direct data participation from 56.96\% to 60.21\% relative to random clustering, but both reach 85\% in similar mean time, $5.70\pm0.50$ and $5.66\pm0.18$ minutes. Label-aware GCPSL reaches 59.43\% participation and 85\% in $6.13\pm0.15$ minutes. Figure~\ref{fig:teaser} shows these operating points.

The factorization in Eq.~\eqref{eq:participation-factorization} places almost the entire participation gain in barrier padding. Size-balanced clustering raises the model-update participation $\mu$ from 94.47\% to 99.86\%. The within-round rate $\bar\pi$ is 60.29\% for random, 60.30\% for size-balanced, and 60.24\% for label-aware clustering. The repartitioned baseline, which reassigns clients every epoch, also has $\bar\pi=60.28\%$. All four rates lie below 63.5\%, the equal-share large-pool value of Proposition~\ref{prop:participation}(b) for clusters of 64 clients. Data-parallel and larger-batch GPSL draw each 256-example update from all 256 clients and reach $\bar\pi=60.11\%$, below their equal-share large-pool value of 63.3\%. Four 64-example workloads thus reach as many clients per round as one fourfold larger batch. In this setting, the assignment rule controls how long clients wait at the barrier, not how many clients each batch reaches.

The same rate sets the gaps between a client's direct contributions. The mean gap is 3.61 inactive system phases for single-worker GPSL and 0.66 to 0.76 for the fixed assignments. These means equal $(1-\pi)/\pi$ to within 0.001. The 95th percentile falls from 13 to 3 phases, and every method reaches every client. The Gini coefficient of per-client contribution counts falls from 0.211 to 0.144--0.146. By Eq.~\eqref{eq:expected-participation}, clients with more data participate more often, so some count inequality is inherent to the sampler.

\subsection{Barrier Savings Are Smaller Than Run-to-Run Variation}
\label{sec:runtime-breakdown}
Table~\ref{tab:runtime-breakdown} reports the training time per epoch and the shares spent at the fusion barrier and in synchronization. Size-balanced clustering shortens the epoch by only 1.9\%, from 14.09 to 13.82 seconds, and lowers the barrier share from 4.1\% to 2.7\%. With $\mu=99.86\%$, the remaining barrier time is consistent with per-rank speed differences.

\begin{table}[t]
\caption{Training time per epoch, synchronization overhead, and epochs to 85\% validation accuracy over three paired runs.}
\label{tab:runtime-breakdown}
\centering\small
\setlength{\tabcolsep}{5pt}
\begin{tabular}{@{}lrrrr@{}}
\toprule
Method & Epoch (s) & Barrier (\%) & Sync.\ (\%) & Epochs \\
\midrule
\tpdsTimingRows
\bottomrule
\end{tabular}
\par\smallskip
\begin{minipage}{0.98\columnwidth}\footnotesize
Epoch is barrier-aligned training time per epoch over 100 epochs, without validation, testing, and checkpoint I/O. Barrier and Sync.\ are rank-averaged shares of training time. Sync.\ is the full epoch-fusion routine, including state checks, or the per-step gradient all-reduce. Synchronized BatchNorm collectives run inside the forward and backward passes and are not included. Epochs to 85\% are mean $\pm$ sample standard deviation.\end{minipage}
\end{table}

Over the 22 epochs to the target, this saves about six seconds. Epochs to 85\% range from 20 to 24 across the size-balanced runs, at about 15 seconds per epoch including validation. The gain is smaller than one epoch of run-to-run variation, so it does not appear in the mean $t_{85}$.

The data-parallel epoch takes 17.87 seconds, 29\% longer than the size-balanced epoch. Data-parallel GPSL spends 4.7\% of rank training time in its per-step gradient all-reduce, compared with 0.4\% for epoch fusion, and its forward and backward passes contain 75 synchronized server BatchNorm collectives per step. It reaches 85\% after 16.7 instead of 22.0 epochs, so the longer epoch and the fewer epochs nearly cancel in its $t_{85}$ of 5.69 minutes.

\subsection{Epoch Fusion Cuts Cross-GPU Synchronization Volume}
GCPSL synchronizes across GPUs only at fusion, whereas the synchronous baselines synchronize gradients and server BatchNorm statistics at every step. Table~\ref{tab:sync} reports the volume of tensors entering these NCCL collectives, summed over the four GPUs, and the measured collective time per GPU over 100 epochs. Each GCPSL variant passes 53.7~GB to fusion collectives, which take 1.0~s per GPU. Data-parallel GPSL passes 3.15~TB, 59 times as much, and spends 195~s in gradient and BatchNorm collectives. Sharded synchronous GPSL takes four times as many steps per epoch and passes 12.6~TB in 580~s. Up to the 85\% target, GCPSL passes 11.8 to 12.3~GB, compared with 526~GB for data-parallel GPSL. The payload in Table~\ref{tab:real-distributed} excludes these collectives for the synchronous baselines. With them, data-parallel GPSL moves 1.64~TB to the target, more than any GCPSL variant.

\begin{table}[t]
\caption{Cross-GPU synchronization on four H100 GPUs, as means over three paired runs.}
\label{tab:sync}
\centering\small
\setlength{\tabcolsep}{4pt}
\begin{tabular}{@{}lrrr@{}}
\toprule
Method & Volume (GB) & To 85\% (GB) & Time (s) \\
\midrule
\tpdsSyncRows
\bottomrule
\end{tabular}
\par\smallskip
\begin{minipage}{0.98\columnwidth}\footnotesize
Volume counts tensors entering fusion, gradient, and BatchNorm collectives over 100 epochs, summed over GPUs. To 85\% counts all collectives up to the target. Time sums the measured durations of the collective calls only, per GPU over 100 epochs.\end{minipage}
\end{table}

%% file: figures/generated/tpds_target_rows.tex
\newcommand{\tpdsTargetRows}{%
Single-worker & 6.31 $\pm$ 1.01 & 10.88 $\pm$ 0.58 & 23.28 $\pm$ 4.53 \\
Random & 3.29 $\pm$ 0.12 & 5.66 $\pm$ 0.18 & 10.07 $\pm$ 0.23 \\
Size-balanced & 3.39 $\pm$ 0.10 & 5.70 $\pm$ 0.50 & 10.20 $\pm$ 0.85 \\
Label-aware & 3.56 $\pm$ 0.15 & 6.13 $\pm$ 0.15 & 11.30 $\pm$ 1.16 \\
Repartitioned & 3.16 $\pm$ 0.13 & 5.84 $\pm$ 0.24 & 10.31 $\pm$ 1.03 \\
Sharded sync. & 4.63 $\pm$ 0.42 & 8.95 $\pm$ 1.49 & 16.85 $\pm$ 4.50 \\
Data-parallel & 2.93 $\pm$ 0.18 & 5.69 $\pm$ 0.48 & 19.53 $\pm$ 3.17 \\
Larger-batch & 5.05 $\pm$ 1.26 & 11.45 $\pm$ 1.42 & 35.45 $\pm$ 11.88 \\%
}

%% file: figures/generated/tpds_timing_rows.tex
\newcommand{\tpdsTimingRows}{%
Random & 14.09 & 4.1 & 0.4 & 21.7 $\pm$ 0.6 \\
Size-balanced & 13.82 & 2.7 & 0.4 & 22.0 $\pm$ 2.0 \\
Label-aware & 13.92 & 2.8 & 0.4 & 22.7 $\pm$ 0.6 \\
Data-parallel & 17.87 & 0.0 & 4.7 & 16.7 $\pm$ 1.5 \\%
}

%% file: figures/generated/tpds_sync_rows.tex
\newcommand{\tpdsSyncRows}{%
Random & 53.7 & 11.8 & 1.0 \\
Size-balanced & 53.7 & 12.0 & 1.0 \\
Label-aware & 53.7 & 12.3 & 1.0 \\
Repartitioned & 53.7 & 12.5 & 1.0 \\
Data-parallel & 3{,}151.7 & 525.5 & 195.2 \\
Sharded sync. & 12{,}606.6 & 1{,}386.9 & 579.7 \\%
}

%% file: figures/generated/v10_4_boundary_tabular.tex
\begin{tabular*}{\columnwidth}{@{\extracolsep{\fill}}lrrrrr@{}}
\toprule
Method & $B$ & $N$ & $R_{50}$ & $R_{60}$ & Acc. (\%) \\
\midrule
GPSL & 128 & -- & 3.39 & 6.52 & 66.66 \\
Label-aware GCPSL & 128 & 4 & 2.26 & 4.05 & 68.31 \\
Size-balanced GCPSL & 128 & 4 & 2.48 & 4.48 & 66.77 \\
\midrule
GPSL & 512 & -- & 0.85 & 1.89 & 64.42 \\
\bottomrule
\end{tabular*}

%% file: figures/generated/v10_4_fusion_tabular.tex
\begin{tabular}{@{}llccc@{}}
\toprule
Execution & Fusion interval & Acc. (\%) $\uparrow$ & $R_{85}$ $\downarrow$ & $R_{88}$ $\downarrow$ \\
\midrule
Concurrent & one epoch & \textbf{89.25} & 3.33 & \textbf{5.90} \\
Sequential & one epoch & \textbf{89.25} & 24.88 & 44.00 \\
Concurrent & 100 steps & 88.15 & \textbf{3.06} & -- \\
Concurrent & none & 28.34 & -- & -- \\
\bottomrule
\end{tabular}

%% file: figures/generated/v10_4_payload_tabular.tex
\begin{tabular}{@{}lllrrr@{}}
\toprule
Method & Clustering & $N$ & GB/phase $\downarrow$ & TB@85 $\downarrow$ & TB@88 $\downarrow$ \\
\midrule
GPSL & -- & -- & \textbf{0.681} & 4.791 & 9.227 \\
GCPSL & label-aware & 4 & 0.869 & 3.059 & \textbf{5.253} \\
GCPSL & size-balanced & 4 & 0.884 & \textbf{3.004} & 5.545 \\
GCPSL & label-aware & 8 & 1.081 & 3.601 & 6.369 \\
GCPSL & size-balanced & 8 & 1.158 & 3.708 & 6.167 \\
\bottomrule
\end{tabular}

%% file: figures/generated/real_results_mean_std_tabular.tex
\begin{tabular*}{\textwidth}{@{\extracolsep{\fill}}>{\raggedright\arraybackslash}p{0.205\textwidth}rrrrrrr@{}}
\toprule
Method & GPUs & $t_{85}$ (min) $\downarrow$ & GPU-min $\downarrow$ & Acc. (\%) $\uparrow$ & Best-val (\%) $\uparrow$ & Part. (\%) $\uparrow$ & Payload (TB) $\downarrow$ \\
\midrule
\multicolumn{8}{@{}l}{\emph{Fixed-assignment comparison}} \\
\addlinespace[1.5pt]
Random & 4 & 5.66 $\pm$ 0.18 & 22.62 & 89.69 & 90.07 & 56.96 & 1.461 \\
Size-balanced & 4 & 5.70 $\pm$ 0.50 & 22.81 & 89.32 & 89.52 & 60.21 & 1.483 \\
Label-aware & 4 & 6.13 $\pm$ 0.15 & 24.51 & 89.53 & 89.73 & 59.43 & 1.528 \\
\midrule
\multicolumn{8}{@{}l}{\emph{Other reference baselines}} \\
\addlinespace[1.5pt]
Single-worker & 1 & 10.88 $\pm$ 0.58 & 10.88 & 89.59 & 89.83 & 21.70 & 1.654 \\
Sequential label-aware & 4 & 19.09 $\pm$ 0.45 & 76.35 & 89.53 & 89.73 & 59.43 & 1.528 \\
Repartitioned periodic & 4 & 5.84 $\pm$ 0.24 & 23.35 & 89.95 & 90.22 & 57.12 & 1.551 \\
Sharded synchronous & 4 & 8.95 $\pm$ 1.49 & 35.78 & 88.79 & 90.36 & 21.70 & 1.705 \\
Data-parallel & 4 & 5.69 $\pm$ 0.48 & 22.77 & 88.21 & 89.43 & 60.11 & 1.114 \\
Larger-batch & 1 & 11.45 $\pm$ 1.42 & 11.45 & 89.08 & 89.58 & 60.11 & 1.136 \\
\bottomrule
\end{tabular*}

%% file: sections/04_discussion_conclusion.tex
\section{Discussion and Limitations}
\label{sec:discussion}

\subsection{Operating Region}
GCPSL targets systems with many stable clients, a fixed per-workload batch, and several server execution slots. The cluster count should follow both the available slots and the ratio $K/B$. The counting bound in Eq.~\eqref{eq:system_idle_bound} vanishes at $NB=K$, but for equal client shares the large-pool participation of Proposition~\ref{prop:participation}(b) keeps rising with $N$, because each batch covers a larger share of its cluster. More clusters also leave each local sampler fewer clients and less label diversity. In the CIFAR-10 sweep, $N=8$ lowers inactivity to 17\% and, with size-balanced clustering, gives the fewest rounds to 88\%, whereas $N=16$ lowers it to 4\% but loses 3.0 to 3.8 accuracy points relative to $N=4$.

The assignment rule should match the available metadata. Random clustering needs no data-derived metadata. Size-balanced clustering needs one sample count per client and nearly removes the padding caused by unequal round counts. On CIFAR-10, this raises participation, but its effect on time to target is smaller than the run-to-run variation. Label-aware clustering also uses class histograms and therefore discloses more about each local dataset. It pays off on CIFAR-100, where it finishes 1.5 percentage points above size-balanced clustering. Fusion is required: without it, label-aware GCPSL at $N=8$ finishes at 28.34\% (Table~\ref{tab:controls}).

GCPSL is preferable when a larger batch does not fit in memory or degrades optimization, and when per-step global synchronization is costly. On four H100 GPUs, random and size-balanced GCPSL matched the fastest alternative at 85\% and were the two fastest methods at 88\%, while synchronizing across clusters once per epoch. A larger single workload remains the simpler choice when its batch fits and its optimization behavior is acceptable.

\subsection{Limitations}
The real-execution study covers one ResNet-18 split, CIFAR-10, 256 logical clients, one four-H100 node, and three paired runs per regime. It does not establish statistical significance or performance for other models, datasets, or allocations. The 256 clients are logical data owners mapped onto four GPU processes, which compute their active clients one after another. Their activations and cut gradients stay inside each process, and only fusion and synchronization collectives cross the node interconnect. Physical client-to-server links, device heterogeneity, intermittent availability, multi-node communication, and slower interconnects would add costs that are absent here, including to the per-step synchronization of the data-parallel baseline, and an optimized BatchNorm synchronization would shorten the synchronous baselines. Client membership is fixed during each GCPSL run.

The schedule-accounted studies measure learning outcomes but count time in schedule rounds without contention. They use no validation split and therefore report test accuracy. Their three runs are independent. Both payload measures count application tensors, not protocol traffic, physical network bytes, or energy. Raw examples remain local, but activations, cut gradients, and model state can leak information \cite{pasquini2021tiger}. GCPSL provides no formal privacy guarantee.

\section{Conclusion}
\label{sec:conclusion}
With many clients and a fixed GPSL batch, most clients cannot contribute data in a round, regardless of server capacity. GCPSL assigns clients to stable clusters, runs one GPSL workload per cluster concurrently, and fuses the split-model replicas at epoch boundaries. The participation analysis shows that, for large client datasets, cluster size limits the expected fraction of a cluster that each batch reaches, while the assignment rule controls barrier padding. On four H100 GPUs, GCPSL raised direct data participation from 22\% to 57--60\%. With random or size-balanced clusters, it reached 85\% validation accuracy 1.6 times faster than sharded synchronous GPSL and as fast as per-step data-parallel GPSL, and it reached 88\% in about half the time of data-parallel GPSL. GCPSL suits systems with stable client assignments and several execution slots that can each run a moderate-batch workload.